\documentclass[aj]{emulateapj}

\usepackage{graphicx}
\usepackage{microtype}
\usepackage{hyperref}  
\def\link_col{blue}
\hypersetup{colorlinks,citecolor=\link_col,filecolor=\link_col,linkcolor=\link_col,urlcolor=\link_col}

\newcommand{\jybeam}{$\textrm{Jy~beam}^{-1}$~}

\slugcomment{The Astronomical Journal, 141:82 (7pp), 2011 March}

\shorttitle{ATCA 1384 and 2368~MHz observations of Sgr~B}
\shortauthors{Jones et al.}

\begin{document}

\title{AUSTRALIA TELESCOPE COMPACT ARRAY RADIO CONTINUUM 1384 AND 2368 MHz OBSERVATIONS OF SAGITTARIUS B}

\author{David I. Jones\altaffilmark{1,2,3}, Roland M. Crocker\altaffilmark{3,4}, J\"{u}rgen Ott\altaffilmark{5,6} Raymond J. Protheroe\altaffilmark{1} and Ron D. Ekers\altaffilmark{2}}
\email{djones@mpi-hd.mpg.de}

\altaffiltext{1}{Department of Physics, School of Chemistry \& Physics, University of Adelaide, South Australia, 5000, Australia.}
\altaffiltext{2}{Australia Telescope National Facility, CSIRO, P.O. BOX 76 Epping, NSW 1710, Australia.}
\altaffiltext{3}{Max Planck Institut f\"{u}r Kernphysik, Postfach 103980, 69029 Heidelberg, Germany.}
\altaffiltext{4}{School of Physics, Monash University, Victoria, Australia.}
\altaffiltext{5}{National Radio Astronomy Observatory, 520 Edgemont Road, Charlottesville, VA 22903, USA.}
\altaffiltext{6}{California Institute of Technology, 1200 E. California Blvd., Caltech Astronomy, 105-24, Pasadena, CA, 91125, USA.}
\begin{abstract}

We present images of the Sagittarius (Sgr)~B giant molecular cloud at 2368 and 1384~MHz obtained using new, multi-configuration Australia Telescope Compact Array (ATCA) observations. We have combined these observations with archival single-dish observations yielding images at resolutions of $47''\times14''$  and $27''\times8''$ at 1384 and 2368~MHz respectively. These observations were motivated by our theoretical work (Protheroe et al. 2008) indicating the possibility that synchrotron emission from secondary electrons and positrons created in hadronic cosmic ray (CR) collisions with the ambient matter of the Sgr B2 cloud could provide a detectable (and possibly linearly polarized) non-thermal radio signal.  We find that the only detectable non-thermal emission from the Sgr~B region is from a strong source to the south of Sgr~B2, which we label Sgr~B2 Southern Complex (SC).  We find Sgr~B2(SC) integrated flux densities of $1.2\pm0.2$~Jy at 1384~MHz and $0.7\pm0.1$~Jy at 2368~MHz for a source of FWHM size at 1384~MHz of $\sim54''$.  Despite its non-thermal nature, the synchrotron emission from this source is unlikely to be dominantly due to secondary electrons and positrons. Failing to find clear evidence of non-thermal emission due to secondary electrons and positrons, we use polarization data to place $5\sigma$ upper limits on the level of polarized intensity from the Sgr~B2 cloud of 3.5 and 3~m\jybeam at 1384 and 2368~MHz respectively. We also use the angular distribution of the total intensity of archival 330~MHz VLA and the total intensity and polarized emission of our new 1384~MHz and 2368~MHz data to constrain the diffusion coefficient for transport of the parent hadronic CRs into the dense core of Sgr~B2 to be no larger than about 1\% of that in the Galactic disk. Finally, we have also used the data to perform a spectral and morphological study of the features of the Sgr~B cloud and compare and contrast these to previous studies.
\end{abstract}


\section{Introduction}\label{sec:intro}
In 2006 the High Energy Stereoscopic System (HESS) gamma-ray telescope discovered a diffuse region of TeV gamma-rays pervading the central regions of our Galaxy ($-1.0^\circ\leq l\leq1.5^\circ$, $|b|\leq0.2^\circ$; \citealt{Aharonian2006}). This diffuse Galactic center (GC) TeV gamma-ray emission shows a good correlation with the molecular material as traced by the CS(1--0) emission line \citep{Tsuboi1999}. The correlation between molecular gas density and diffuse gamma-ray intensity suggests that the gamma-rays originate in collisions between the in-situ, hadronic CR population and the ambient gas. Given, however, the total mass of molecular gas in the GC, the measured flux of gamma-rays implies that the GC CR flux at 10~TeV is $\sim10$ times that observed at the top of the Earth's atmosphere. The diffuse TeV emission is particularly bright around Sgr B suggesting that there could be an associated and measurable non-thermal $\sim$GHz flux density from the region due to synchrotron emission from secondary electrons and positrons. These are created from the decay of charged pions (themselves created in $pp$ collisions), $\pi^\pm\rightarrow\mu^\pm\rightarrow e^\pm$, the latter produced concomitantly with the neutral pions postulated to supply the observed TeV gamma-rays via $\pi^0\rightarrow\gamma\gamma$ \citep{Crocker2007}.

In a previous paper, Protheroe et~al. (2008) (hereafter \cite{Protheroe2008}), we have shown that GHz synchrotron emission from secondary electrons and positrons created in the Sgr~B2 cloud  could possibly provide a detectable, non-thermal radio signal. Our modeling also indicated that there could be a relative dimming of the center of the cloud due to the probable exclusion of multi-GeV CRs from the densest parts of the cloud (leading to a `limb-brightened' morphology). We have, therefore, undertaken a series of observations of the Sgr~B giant molecular cloud (GMC) complex using the ATCA at 1.4 and 2.4~GHz in order to compliment the lower frequency, 330~MHz VLA observations of \cite{LaRosa2000} and search for non-thermal radio emission from the Sgr~B2 GMC. The ATCA -- at least until the new Extended VLA (EVLA) 2.4~GHz receiver comes online -- is the only aperture synthesis radio telescope capable of observing Sgr~B with arcsecond resolution at 2.4~GHz ($\lambda13$~cm), but there have previously existed no sensitive, sub-arcminute resolution images of Sgr~B at $\sim2.4$~GHz.  We rectify this deficit here. The wide field of view (FoV) of ATCA at 2368~MHz provides, in addition, an excellent opportunity to fill a gap in the Sgr B spectrum  of 3.4~GHz in frequency (from 1.4 to 4.8~GHz, or almost 2 octaves). 

The new 1384 and 2368~MHz observations are described in \autoref{sec:obs} along with details of the archival data and image deconvolution.  New images of the Sgr~B region at 1384 and 2368~MHz are then presented in \autoref{sec:results} in addition to a detailed discussion of the morphology of the sources observed and a comparison to published work. In \autoref{sec:Synch} we describe how we place upper limits on the flux density of possible non-thermal emission from Sgr B2 and turn this into a constraint on the production of secondary electrons in the cloud. This upper limit, in turn, implies a suppression of the diffusion of multi-GeV CRs into the dense cloud cores found within Sgr~B2.  Our conclusions are presented in \autoref{sec:conclusions}.  

\begin{figure}[h]
\centering
\includegraphics[width=0.4\textwidth,angle=90]{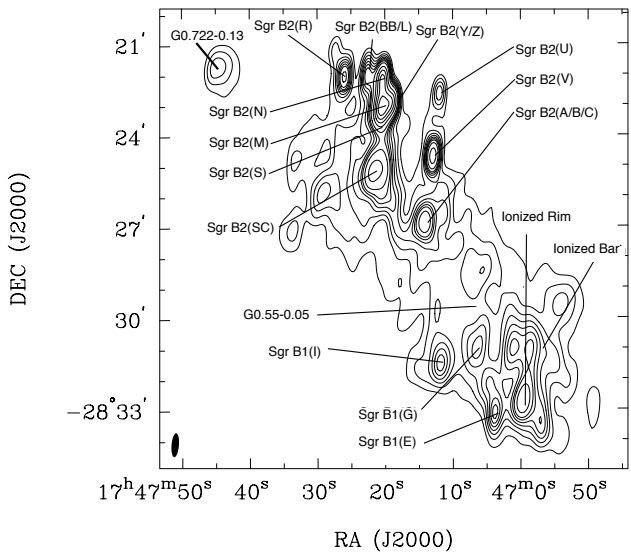}
\caption{Image of 1384~MHz total intensity contours of the Sgr~B region showing the sources as labeled in \autoref{table:SgrBSPIN}.  Contours are at 200, 250, 300, 350, 400, 450, 500, 600, 800, 1000, and  1600 m\jybeam.  The image has a resolution of $47''\times14''$, which is illustrated by the beam in the lower left-hand corner.\label{sgrB-guide}}
\end{figure}

Throughout this paper we have attempted, for simplicity, to follow the naming convention from \citet{Mehringer1992,Mehringer1993}, which is illustrated in \autoref{sgrB-guide}.  Due to the large discrepancy in resolution between our ATCA observations and those of Mehringer et al. (\citealt{Mehringer1992,Mehringer1993}), however, some sources in our images are inevitably confused.  In these cases, we have agglomerated the names of the sources into a single source name.  For this reason, flux density comparisons are not possible between the images presented here and the Mehringer et al. data.

\section{Radio Continuum Observations: New and Old}\label{sec:obs}

\subsection{New ATCA observations }
\begin{figure}[h]
\centering
\includegraphics[height=0.5\textwidth,angle=90]{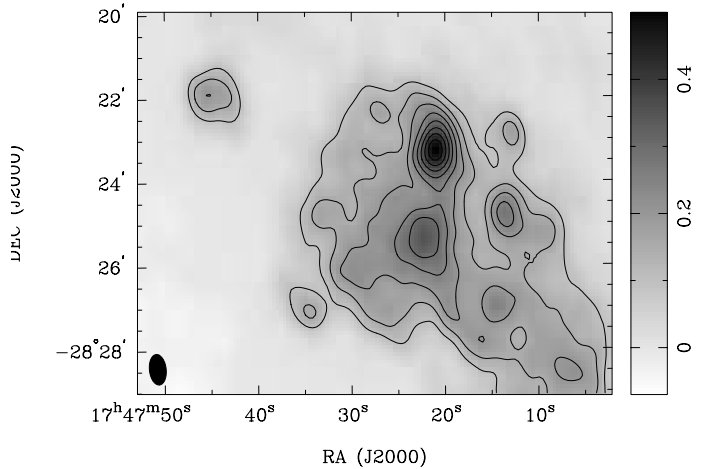}
\caption{330~MHz total intensity medium resolution image of Sgr B2 with a beam of $43''\times24''$ (located in the lower left corner; \citealt{LaRosa2000}).  The contours are 330~MHz total intensity contours at 0.1, 0.15, 0.2, 0.25, 0.3, 0.35, 0.4 and 0.45 \jybeam.  The intensity scale on the right is in units of \jybeam and is set from -0.07 to 0.5 Jy.\label{330MHz-medRes}}
\end{figure}

\autoref{table:ATCAobs} lists the details of the observations taken with the ATCA, which are described in detail below. We chose the ATCA configurations to match the resolution of the 330~MHz VLA observations of \cite{LaRosa2000}. This data, as shown in \autoref{330MHz-medRes} has a resolution of $43''\times24''$ and an RMS sensitivity of $\sim5$~mJy~beam$^{-1}$.

\begin{deluxetable}{ccccc}
\tablecaption{ATCA Observational parameters\label{table:ATCAobs}}
\tablewidth{0pt}
\tablehead{
\colhead{$\nu$ (MHz)} & \colhead{$\Delta\nu$ (MHz)} & \colhead{Array} & \colhead{Time (hours)} & \colhead{Date} }
\startdata
	1384, 2368  & 128 & 1.5C \tablenotemark{a} & 10 & 2005 Dec \\
	1384, 2368 & 128 & 750A \tablenotemark{b} & 10 & 2007 Jan\\
	1384, 2368 & 128 & 750D \tablenotemark{b}   & 8 &  2007 Mar \\
\enddata
\tablenotetext{a}{Single pointing observation.}
\tablenotetext{b}{7 pointing, hexagonal Nyquist-sampled observation.}
\end{deluxetable}

\subsubsection{1.5C array configuration}\label{subsec:1500Config}
Single pointing ATCA radio continuum observations were conducted on 2005 December 4th using the 1.5~C array-configuration utilizing five of a possible six antennae and covering baselines on an east-west track from 76.5\,m to 1.4\,km.  Observations were conducted simultaneously at 1384~MHz ($\lambda$20\,cm) and 2368~MHz ($\lambda$13\,cm), each spanning 128~MHz in bandwidth and recording all four polarization products.  The FWHM primary beam (FoV) of the ATCA is 34$'$ at 1384~MHz and 20$'$ at 2368~MHz.  The total integration time was 600 minutes obtained in 45 minute cycles for approximately 10 hours.  Flux density calibrations were undertaken with PKS~1934-638 assuming flux densities of 15 and 11.6~Jy at 1384 and 2368~MHz respectively.  Phase and gain calibrations were undertaken using PKS~1740-517 from the ATCA calibrator catalog.  

\subsubsection{750m array-configurations}
We have performed a 7-pointing mosaic observation of the entire Sgr B region using the 750A and 750D array-configurations of the ATCA.  The addition of the mosaic data gives the images sensitivity to larger scale emission which the addition of the single-dish data cannot.  The combination of the two 750 array-configurations (750A and 750D) has east-west baselines from 30m to 750m in 20m increments.  We used the same correlator configuration as for the 1.5C array configuration observations with 128~MHz bandwidth with frequencies centered at 1384 and 2368~MHz recording all four polarization products.  We followed a fully-sampled, hexagonal pointing mosaic centered, in J2000.0 coordinates, at a Right Ascension and Declination of $(\alpha,\delta)$=(17:47:07.13, -28:29:32.14) for a total coverage of $\sim1^\circ$ and providing uniform sensitivity over the entire Sgr B region.  The flux, and phase and gain calibrators were PKS 1934-638 and PKS 1740-517 respectively.

\subsection{Archival single-dish data}\label{sec:SD}
In order to fill in the missing short-spacing information for the new ATCA images, we have obtained archival single-dish data at 1408 and 2400~MHz from wide-field surveys obtained using the Effelsberg (EBG; \citealt{Reich1986}) and Parkes (PKS; \citealt{Duncan1995}) telescopes respectively.

The 1.4~GHz data comes from the EBG wide-field survey at 1420~MHz, covering the sky within a Declination range of $-19^\circ<\delta<90^\circ$, with a resolution of $9'.1$\footnote{available from: \url{http://www.mpifr-bonn.mpg.de/survey.html}}.  The RMS noise is 125~m\jybeam with an estimated calibration error of $\sim5$\%.  The 2.4~GHz PKS Galactic plane survey covered $\sim40^\circ$ in Galactic longitude, and $\pm5^\circ$ in Galactic latitude and includes the GC at a resolution of $\sim10'$.  The RMS noise of the survey is $\sim12$~m\jybeam with a bandwidth of 145~MHz.

\subsubsection{Image deconvolution}
To produce the total intensity images presented in this paper, each set of interferometer data was calibrated separately using standard calibration procedures of the MIRIAD software package.  Imaging was then performed by Fourier transforming the entirety of the interferometer data using the MIRAD task \emph{invert}, and deconvolved using the MIRIAD task \emph{mosmem}.  We have also produced images of the Stokes Q, U and V parameters at both 1384 and 2368~MHz (without the addition of single-dish data).  The data were calibrated using the standard MIRIAD data calibration described above.  To produce the images, we used the MIRIAD task {\it pmosmem}, which performs a joint maximum entropy deconvolution of the total and polarized intensities for mosaic observations simultaneously \citep{Sault1999}.  A maximum entropy algorithm was preferred (to a CLEAN-based method) because it recovers large scale structures measured by the mosaicing process \citep{Ekers1979} and allows the Stokes Q, U and V images to have negative values.  Images of linearly polarized intensity, $L=(Q^2+U^2)^{1/2}$, linearly polarized position angle, $\theta=1/2\tan(U/Q)$ and fractional polarization, $L/I$, where $I$ is the total intensity were produced using the standard MIRIAD task, {\it impol}, blanking the output below a level of $5\sigma$. Following deconvolution, each image was smoothed with a Gaussian restoring beam of FWHM of $47''\times14''$ and $27''\times8''$ at 1384 and 2368 MHz respectively and have an RMS noise limit of 10~m\jybeam at 1384~MHz and 5~m\jybeam at 2368~MHz.  

The single-dish data at 1.4 and 2.4~GHz from \cite{Duncan1995} and \cite{Reich1986} described in \autoref{sec:SD} were then combined with the interferometer data using the MIRIAD task {\it immerge}.  The single-dish images used were of a much larger field ($\sim10^\circ\times5^\circ$; \citealt{Crocker2010}), centered on the GC which were re-gridded using the MIRIAD task {\it regrid} onto a common grid with the interferometric data. The data were then combined after image deconvolution of the interferometer data \citep{Stanimirovic2002}.  This `linear' method works on the assumption that the single-dish image is a good representation of the object at low spatial frequencies, whereas the interferometer data better represents the high spatial frequencies.  The image is created by tapering (in the Fourier domain) the low spatial frequencies of the interferometer data so that the sum of the single-dish and interferometer (tapered) data is a gaussian beam equal to the beam of the interferometric data.  The resulting images have the same resolution as the interferometer data.

\section{Results \& discussion}\label{sec:results}
In this section, we present images of the Sgr~B region produced using the data described in the previous section and discuss the spectral and morphological features of the region.  

\subsection{Sgr B morphology}
\begin{figure*}[t]
\centering
\includegraphics[height=0.5\textwidth,angle=90]{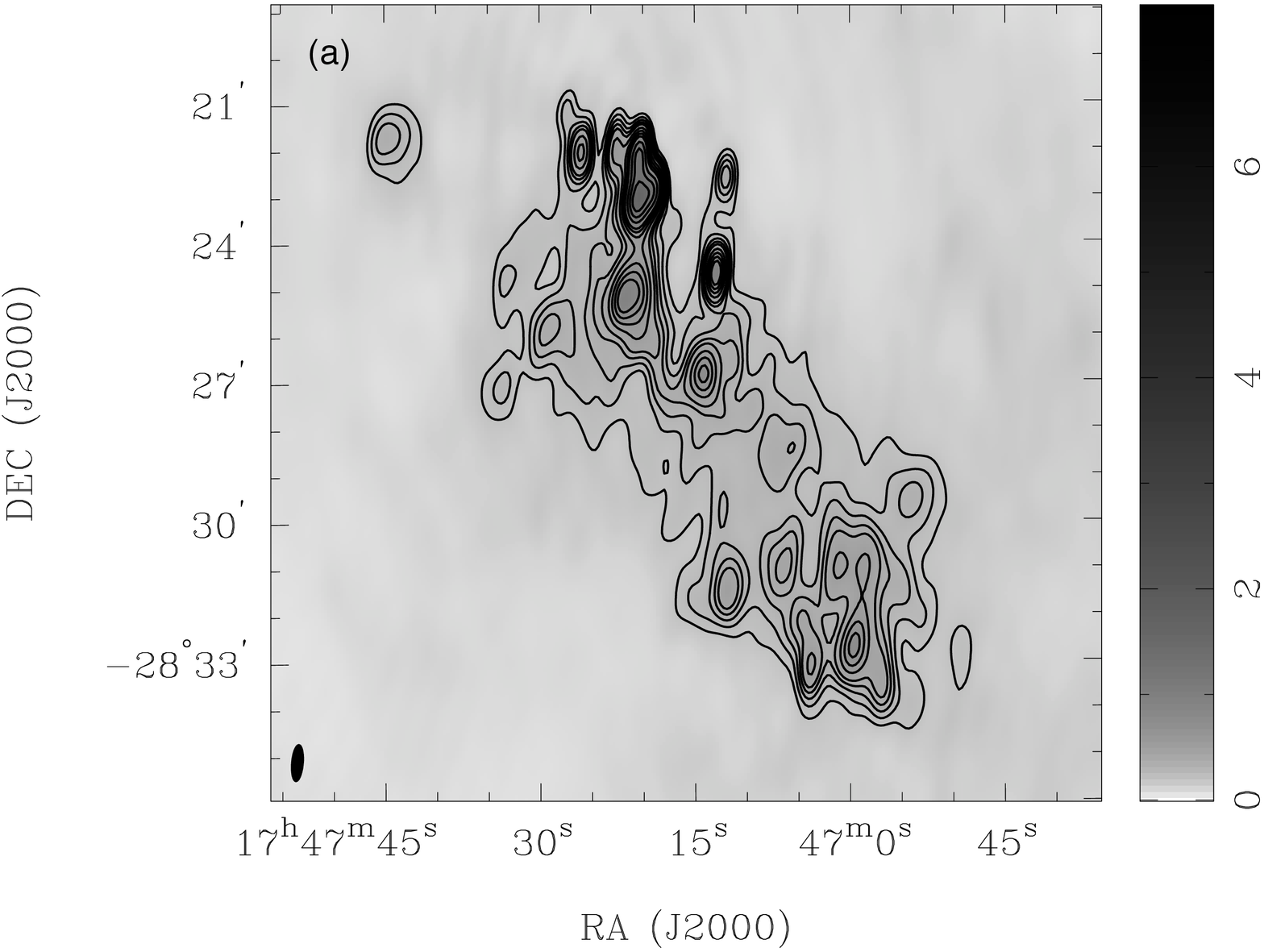}\includegraphics[height=0.5\textwidth,angle=90]{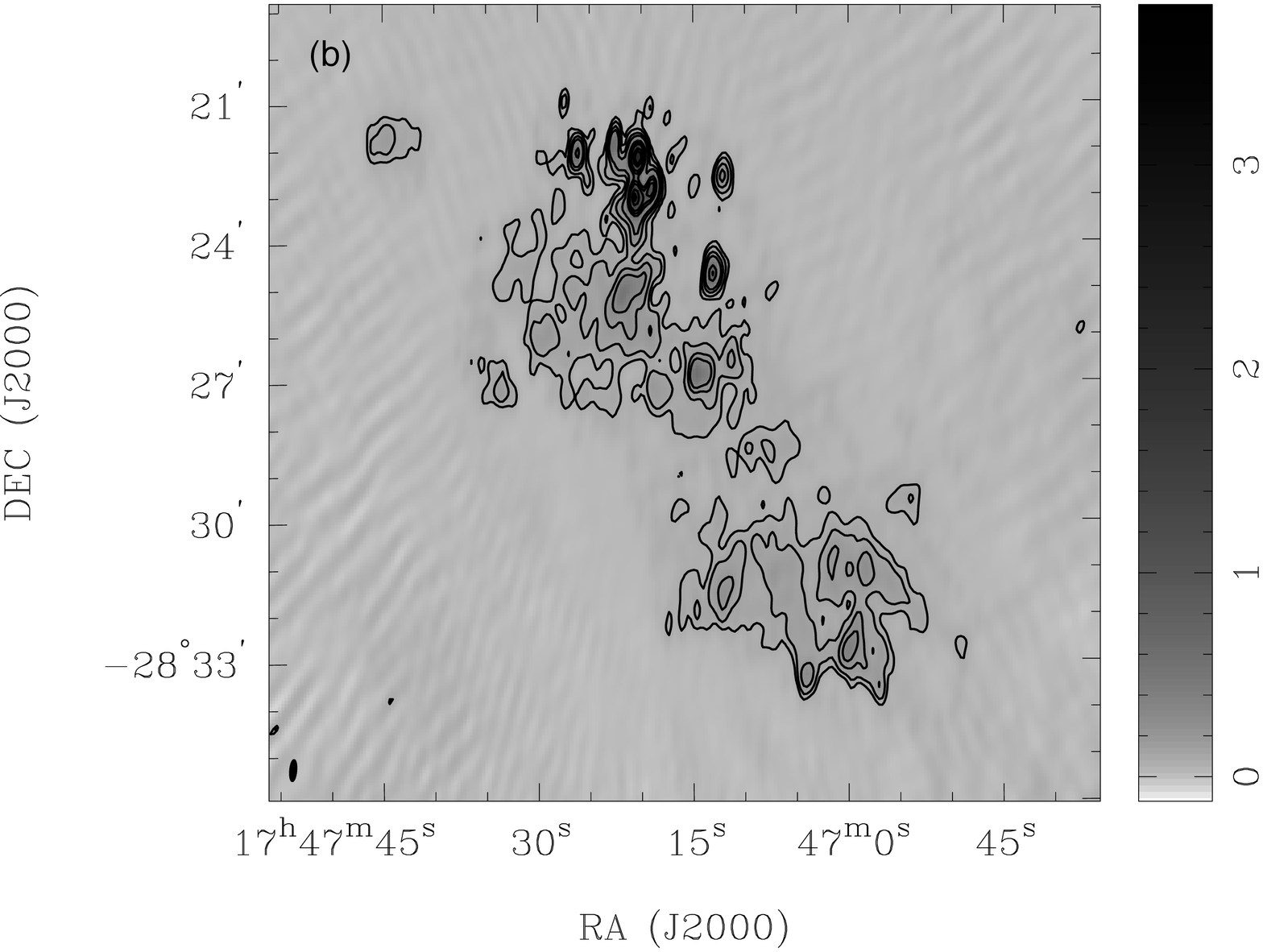}
\caption{Closeup of the Sgr B region at (a) 1384~MHz and (b) 2368~MHz.  The 1384~MHz image has contours at: 0.2, 0.25, 0.3, 0.35, 0.4, 0.5, 0.6, 0.7, 0.8, 1.0, 1.2, and 1.4 \jybeam and a resolution of $47''\times14''$ with the beam located in the lower left hand corner.  The 2368~MHz image has contours at: 0.08, 0.12, 0.2, 0.3, 0.64, 0.8, 1.0, 1.2, 1.5, and 1.8 \jybeam and a resolution of $27''\times8''$ with the beam located in the lower left hand corner.  The RMS noise level of the images are 10~m\jybeam at 1384~MHz and 5~m\jybeam at 2368~MHz, matching the sensitivity at 1384~MHz of similar resolution images.}
\label{completeDataSgrB}
\end{figure*}

At radio continuum frequencies, the Sgr B complex is the second-brightest source in the GC region, with only Sgr~A being brighter. \autoref{completeDataSgrB}(a) and (b) shows that the Sgr~B region is composed of two sub-regions; Sgr~B1 (G0.5-0.0) and Sgr~B2 (G0.7-0.0) and, assuming a distance to the Galactic center of 8.5~kpc, lies at a projected distance of $\sim100$~pc from the GC. Both Sgr~B1 and B2 have been well studied at all wavelengths, including centimeter radio continuum and exhibit a rich chemistry, containing many maser lines and a complex kinematical structure \citep{Jones2008}; Sgr~B1 is situated to the southwest of -- and possibly in front of \citet{Bieging1980} -- Sgr~B2. Sgr B1 is dominated by diffuse H{\sc ii} regions and weak maser activity, whereas Sgr~B2 has many (upwards of 60) compact and ultra-compact H{\sc ii} regions; \citealt{Gaume1995}), implying that Sgr B1 is more evolved than Sgr B2 (\cite{Mehringer1995} and references therein).  \autoref{table:SgrBSPIN} shows that the spectral indices of sources in Sgr~B between 1384 and 2368~MHz are different: Sgr B2 contains sources which have, on average, a larger spectral index (i.e., more postitive where the spectral index, $\alpha$, is defined as $F_\nu\propto\nu^\alpha$, where $F_\nu$ is the flux density at frequency $\nu$) than those found in Sgr~B1 and is consistent, spectrally, with the previous observations. Sgr~B also lies upon a broad region of Galactic plane emission which is particularly evident at 1384~MHz (\autoref{completeDataSgrB}a), but largely absent at 2368~MHz  (\autoref{completeDataSgrB}b). 

The complex nature of this source makes a direct study of the spectral index in the region difficult.  \cite{Law2008}, using a background subtraction technique, found most of the emission from the Sgr~B region to be thermal at 4.85 and 8.5~GHz, except for two positions, one to the east of Sgr~B2 and one between Sgr~B2 and B1.  Interestingly, the latter non-thermal source, at $(l,b)=(0.61,-0.01)$, was also found by \citep{Koyama2007} to exhibit an extended morphology in 6.4~keV thermal line emission and interpreted by them to be a young SNR; we do not, however, observe any non-thermal emission at this position (c.f. \autoref{sec:threeSources}). It should be noted that the X-ray Fe-line emission from the Sgr~B region suggests that the origin of the diffuse X-ray emission. There is also compact X-ray emission from Sgr B which arises from massive stars \citep{Mauerhan2010}; this emission is interpreted to be an X-ray reflection nebula \citep{Koyama2007}. \cite{YusefZadeh2007}, on the other hand, have claimed the presence of non-thermal emission due to an excess in the measured spectrum at low-frequencies from Giant Meter-wave Radio Telescope (GMRT) and VLA observations.  All these examinations, to a certain extent, rely on the fact that non-thermal emission exhibits a negative spectral index (i.e., a spectral index of approximately $\alpha=-0.7$), whereas thermal emission has a spectral index of approximately $\alpha=0.0$ for optically thin thermal emission and approximately $\alpha=+2$ for optically thick thermal emission (``optically thick'' defines emission for which the optical depth, $\tau_\nu$, is much greater than unity, as determined by the observation of absorption, emission or molecular lines).  Such a clear trichotomy of spectral indices is something which, in an environment as complex as Sgr~B, is difficult to ascertain precisely.  This point is underscored by the spectral indices listed in \autoref{table:SgrBSPIN}, where sources exhibit a range of spectral indices, indicating an environment where all of non-thermal, optically thin and thick thermal emission {\it may} be contributing at some level. 

We measure a flux density from the entire region of $\sim70\pm4$ Jy at 1384~MHz and $\sim80\pm4$ Jy at 2368~MHz.  The 1384~MHz flux density agrees well with the 1.4~GHz flux density of $83\pm7$~Jy  from a similar solid angle stated in \cite{Lang2008}.  The spectral index derived from our flux density measurements is also consistent with that obtained by \cite{Law2008} between 4.8 and 8.5~GHz using data obtained with the Green Bank Telescope (GBT).  We conclude then, that the 2.4~GHz flux density is consistent with the conclusions of \cite{Lang2008} that the integrated flux from the Sgr~B region is thermal in nature. This result does not contradict the previous claim of non-thermal emission from this region by \cite{Crocker2007} who measured the spectrum over the region used for analysis by the HESS gamma-ray telescope by \cite{Aharonian2006}. The region presented in \cite{Crocker2007} is much larger (about 0.96 square degrees) so that the non-thermal emission reported there is probably due to the diffuse non-thermal (DNS) source reported by \cite{LaRosa2005}.

\subsubsection{Sgr B1}
\autoref{completeDataSgrB} shows that Sgr~B1 consists of $\sim5$ sources which are labeled as Sgr B1(E,I,G), the Ionized Rim and the Ionized Bar in \autoref{table:SgrBSPIN} and \autoref{sgrB-guide}.  The Ionized Bar is the source to the northwest of Sgr~B1 in \autoref{completeDataSgrB} at 2368~MHz, which is not fully resolved at 1384~MHz.  To the south of this is the Ionized Rim and Source~E -- which shows an extension to the north consistent with Ridge~2 from \cite{Mehringer1992}.  The source to the northeast is Source~I.  Source confusion means that the flux densities presented in \autoref{table:SgrBSPIN} should be taken as a rough guide only -- higher resolution images at 2368~MHz are needed to properly categorise the emission from these regions at these frequencies. 

We find a total flux density of $27\pm1.4$~Jy at 1384~MHz, which is consistent with the 1.4~GHz flux density found by \cite{Mehringer1992}.  Due to the addition of the low spatial frequencies contained in the single-dish data, a component of this flux density will be due to foreground/background diffuse emission.  This contribution to the total flux density at both frequencies was derived in the following way: the average total flux density of a number of similar-sized regions was derived and subtracted from the total flux density stated above.  Thus, for the entire Sgr~B1 region {\it only} (i.e., excluding fore/backgrounds), we find a flux density of $14.8\pm0.8$~Jy at 1384~MHz and $12.8\pm0.6$~Jy at 2368~MHz, broadly consistent with optically thin emission from H{\sc ii} regions.  

\begin{deluxetable*}{cccccccc}
\tabletypesize{\scriptsize}
\tablecaption{Discrete sources within Sgr B}
\tablewidth{\textwidth}
\tablehead{
    \colhead{Name} &  \colhead{R.A.} &  \colhead{Dec.} &  \colhead{$I_{1.4}$} \tablenotemark{a} &  \colhead{$I_{2.4}$}\tablenotemark{a} &  \colhead{S$_{1.4}$} &  \colhead{$S_{2.4}$} &  \colhead{$\alpha^{1384}_{2368}$}\tablenotemark{b} \\
    &  (J2000) & (J2000) & (Jy/Beam) & (Jy/Beam) & (Jy) & (Jy) & \\
    (1) & (2) & (3) & (4) & (5) & (6) & (7) & (8) \\
    }
\startdata
	Sgr B2(M) & 17:47:20.5 & -28:22:53.7 & $2.0\pm0.1$ & $1.2\pm0.1$ & $3.4\pm0.2$ & $3.8\pm0.2$ & $0.2\pm0.3$\\
	Sgr B2(N) & 17:47:20.1 & -28:22:18.3 &       $1.8\pm0.1$ &    $1.3\pm0.1$ &  $2.5\pm0.1$ & $3.5\pm0.2$ &  $0.6\pm0.3$  \\
	Sgr B2(S) \tablenotemark{c} & 17:47:20.5 & -28:23:34.3 & $1.5\pm0.1$ & $0.8\pm0.1$ & $1.5\pm0.1$  & $2.5\pm0.1$ & $1.0\pm0.3$ \\
	Sgr B2(BB/L)\tablenotemark{c} &17:47:22.4 &    -28:21:57.8   & $0.5\pm0.1$ & $0.3\pm0.1$  & $1.2\pm0.1$ & $1.1\pm0.1$ & $-0.2\pm$0.3 \\
	Sgr B2(Y/Z)\tablenotemark{c} &17:47:18.6 & -28:22:54.0 & $1.0\pm0.1$ & $0.5\pm0.1$ & $1.6\pm0.1$ & $1.8\pm0.1$ & $0.2\pm0.3$ \\
	Sgr B2(R) & 17:47:25.8 & -28:22:05.4 & $0.6\pm0.1$ & $1.0\pm0.1$ & $1.4\pm0.1$ & $1.7\pm0.1$ & $0.4\pm0.3$ \\
	Sgr B2(SC) & 17:47:21.7 & -28:25:20.4 & $1.0\pm0.1$ & $0.5\pm0.1$ & $1.2\pm0.1$ & $0.7\pm0.1$ & $-0.9\pm0.3$ \\
	Sgr B2(A/B/C) & 17:47:14.1 & -28:26:52.3 & $0.7\pm0.1$ & $0.4\pm0.1$ & $1.2\pm0.1$ & $1.1\pm0.1$ & $-0.2\pm0.3$ \\
	Sgr B2(V) & 17:47:12.9 & -28:24:35.9 & $0.8\pm0.1$ & $0.6\pm0.1$ & $1.1\pm0.1$ & $1.2\pm0.1$ & $0.2\pm0.3$\\
	Sgr B2(U)\tablenotemark{d} & 17:47:11.8 & -28:22:4.54 & $0.4\pm0.1$ & $0.5\pm0.1$ & $0.4\pm0.1$ & $0.5\pm0.1$ & $0.4\pm0.4$\\
	Ionized Rim\tablenotemark{d} & 17:46:59.5 & -28:32:43.0 & $0.8\pm0.1$ & $0.9\pm0.1$ & $0.8\pm0.1$ & $0.9\pm0.1$ & $0.3\pm0.3$ \\
	Sgr B1(I) & 17:47:11.9 & -28:31:29.3 & $0.5\pm0.1$ & $0.6\pm0.1$ & $2.4\pm0.2$ & $2.7\pm0.2$ & $0.2\pm0.3$ \\
	Sgr B1(G) & 17:47:06.2 & -28:31:01.6 & $0.4\pm0.1$ & $0.5\pm0.1$ & $1.6\pm0.1$ & $1.7\pm0.1$ & $0.1\pm0.3$ \\
	Sgr B1(E) & 17:47:03.8 & -28:33:06.9 & $0.5\pm0.1$ & $0.6\pm0.1$ & $1.8\pm0.1$ & $2.1\pm0.1$ & $0.3\pm0.3$ \\
	Ionized bar\tablenotemark{c} & 17:46:58.6 & -28:30:58.2 & $0.5\pm0.1$ & $0.6\pm0.1$ & $5.6\pm0.3$ & $5.3\pm0.3$ & $-0.1\pm0.3$ \\
\enddata
\tablecomments{\label{table:SgrBSPIN}Units of right ascension are hours, minutes, seconds, and units of declination are degrees, arcminutes and arcseconds (J2000).}
\tablenotetext{a}{This is the peak intensity of the source in units of Jy beam$^{-1}$. The errors presented here are mainly due to calibration errors at the 5\% level and were obtained using the MIRIAD task {\it gpplt}.}
\tablenotetext{b}{The spectral index error is a 5-$\sigma$ error.}
\tablenotetext{c}{At the resolution of the 1384~MHz images these sources are confused by other sources.}
\tablenotetext{d}{Denotes point sources, where the integrated flux densities are the peak flux densities.}
\end{deluxetable*}

\subsubsection{Sgr B2}\label{sec:SgrB2}
\autoref{completeDataSgrB} shows that Sgr~B2 is the brightest part of the Sgr~B region and consists of the compact cores of Sgr~B2(M) and (N), with an extension to Sgr B2(M) in the south, which can be observed at 2368~MHz and is labelled Sgr B2(S).  At 1384~MHz these features are only partially resolved.  The three sources -- Sgr~B2(M) and (N) in particular -- are known to be very active star-forming regions on the basis of the chemistry and numerous compact H{\sc ii} regions (\cite{Jones2008} and references therein).

At 2368~MHz, the smaller beam size also partially resolves two sources: one to the east and the other to the west of Sgr B2(N).  Both sources are known H{\sc ii} regions, Sgr B2(Y/Z) (G0.666-0.03) and Sgr~B2(BB/L) (G0.689-0.03).  These are located to the west and east, respectively, of Sgr B2(M) and (N), and are physically associated with the Sgr B2 cloud \citep{Mehringer1995}.

\subsubsection{Sgr B2(A/B/C), (V) and (U)}\label{sec:threeSources}
Sgr B2(A/B/C) is coincident with the IRAS source IRAS17440-2825, located $1.5''$ away from the peak of the radio emission \cite{Valtts1999}, which contains a Class {\sc ii} methanol maser source \citep{Slysh1999} and seems to be associated with Sgr~B2 \citep{Mehringer1995}.  At a distance of $4.32''$ from the radio continuum peak presented here is the radio source DGSW 60 \citep{Downes1979}.  This source has a flux density of 0.8~Jy at 5~GHz, spectrally consistent with an optically thin H{\sc ii} region between 1384 and 5000~MHz and the flux densities presented in \autoref{table:SgrBSPIN}. 

\autoref{table:SgrBSPIN} shows that Sgr B2(V) is an H{\sc ii} region which is optically thin between 1384 and 2368~MHz.  This source is listed as MD72(62) by \cite{Downes1979}, $2.3''$ distant from the peak in our images. They find flux densities of 1.4 and 1.1~Jy at 5 and 10.4~GHz respectively -- which implies that our flux density measurements of 1.1 and 1.2~Jy at 1384 and 2368~MHz are consistent with their classification of this source.

Sgr B2(U) is also detected in \cite{Downes1979} and has a flux density of 0.6~Jy at 5~GHz, although there is a difference in central position of $5.4''$.  Our flux density measurements of 0.4 and 0.5~Jy at 1384 and 2368~MHz, respectively, is consistent with an optically thin H{\sc ii} region between 1384 and 5000~MHz. We suggest that although there is an offset of $5.4''$ between \cite{Downes1979} and our data, this is the same source.

\subsubsection{Sgr B2(SC) (G0.637-0.06)}\label{sec:SgrB2SC}
There is, to the south of the main condensates of Sgr~B2, a source which our observations find to be non-thermal, with a spectral index between 1384 and 2368~MHz of $\alpha=-0.9\pm0.3$.  The non-thermal nature of this source has been suggested before by \cite{Gray1994}, and is described in \cite{Protheroe2008} as the `Non-Thermal Source'.  \cite{LaRosa2000} list this source in Table~2 (source 62) of their paper with a flux density of 660~mJy for a source size (angular FWHM) of $54''$, which agrees well with the source size observed in \autoref{completeDataSgrB} so that, compared to our flux density estimates, this source would seem to be absorbed at 330~MHz.  Indeed, comparing the flux density estimates from \cite{LaRosa2000} of $14.01\pm0.69$~Jy and $17.42\pm0.81$~Jy for Sgr~B1 and B2, respectively, suggests that the whole Sgr~B complex could be significantly absorbed by thermal gas at 330~MHz (i.e., $\tau_{330}\gg1$) -- a conclusion which is suggested in \cite{Law2008}. 

We have obtained a theoretical estimate of the synchrotron flux density due to secondary electrons for Sgr~B2(SC).  In this calculation we use: a magnetic field estimate of 0.8~mG (as described in \cite{Protheroe2008} and based on the Zeeman splitting observation of H~{\sc i} line from Sgr~B2 which probably apply to the envelope of the Sgr~B2 cloud, and not the dense cores; \citealt{Crutcher1996}); a number density of $10^3$~cm$^{-3}$ obtained from the NH$_3$ observations of \cite{Ott2006}; and a (linear) size of Sgr~B2(SC) -- FWHM at 8.5~kpc of $\sim2.5$~pc. We obtain at most $\sim1$~mJy of radio continuum flux from Sgr~B2(SC) dominated by synchrotron emission from secondary electrons at 1~GHz (this can be increased to 10~mJy by over-normalizing the CR flux by a factor of 10; cf. \autoref{sec:intro}). 

This result is not surprising since, for a high enough ambient gas density, the synchrotron emissivity is a function of the {\it volume}, not the density: for a fixed magnetic field and volume, the increased injection rate of secondary particles by increasing the density is almost exactly cancelled by the increased efficiency of the bremsstrahlung cooling, leaving the steady state distribution unchanged.  Explicitly, this is because the flux density of synchrotron from secondary electrons is $F_{synch}(2e) \propto N_{2e}$, where $N_{2e}$ is the number of secondary electrons, and: $N_{2e} \equiv n_{2e} V\simeq t_{cool}(e) \dot{q_{2e}} V$.  Here, $\dot{q_{2e}} \propto n_H n_{CR}$ is the production rate of secondary particles, $t_{cool}(e) \simeq t_{cool}(e)^{brems} \propto1/n_H$ is the cooling time of those particles, and $V$ is the source volume. These relations demonstrate that $N_{2e} \propto n_{CR} V$. We conclude, then, that the non-thermal emission from the Sgr~B2(SC) is not due to synchrotron emission from secondary electrons.

\section{Limits on the diffusion of CRs into the Sgr~B2 cloud}\label{sec:Synch}
As we did not uncover any non-thermal emission from the Sgr~B2 cloud (including evidence for limb-brightening) except from Sgr~B2(SC) (see \cite{Jones2009} for a more detailed discussion), we now use the angular distribution of the total intensity and polarization data to place upper limits on the flux density of non-thermal emission from the Sgr~B2 cloud.  These upper limits then imply a constraint on the diffusion of multi-GeV hadronic CRs into the Sgr~B2 cloud.

We follow \cite{Gabici2007} in adopting an energy dependent diffusion coefficient, $D(E)$, of the form (normalized to the Galactic plane value):
\begin{equation}\label{eq:diff}
	D(E)=3\times10^{27}\chi\left[\frac{E/(1\textrm{~GeV)}}{B/(3\textrm{~}\mu\textrm{G)}}\right]^{0.5}\textrm{~cm}^2\textrm{~s}^{-1},
\end{equation}
where $B$ is the magnetic field amplitude, $E$ is the energy, and $\chi$ is a factor to account for the possible suppression (slowing) of diffusive transport into the cloud. Given the absence of detectable non-thermal emission from Sgr~B2, we can place limits on the diffusion of CRs into the Sgr~B2 cloud by requiring that the predicted angular intensity of synchrotron emission from secondary electrons does {\it not} exceed the observed angular intensity distribution.

\subsection{Total intensity limits}
\begin{figure}[h]
\centering
\includegraphics[height=0.5\textwidth,angle=90]{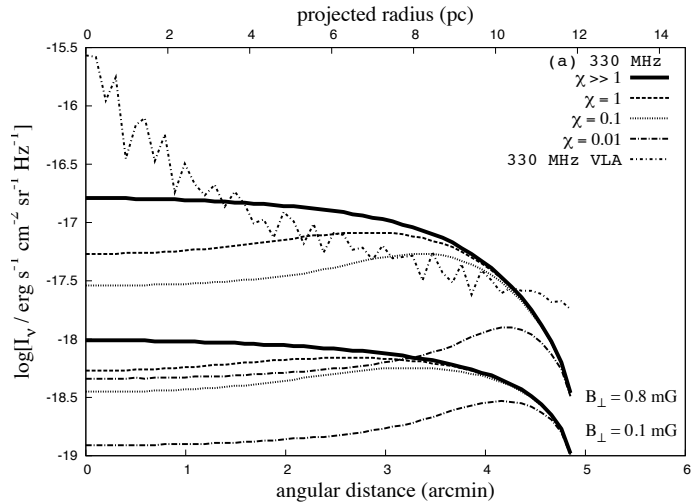}
\caption{Comparison of the actual emission intensity as a function of angular distance from the center of Sgr~B2 and that expected from models of synchrotron emission from secondary electrons at 330~MHz.  The curves are for the parameters labelled and match those of Figure~9 of \cite{Protheroe2008}.\label{fig:lbul}}
\end{figure}

We have used the MIRIAD task {\it ellint}, which finds the radial brightness distribution, or flux density as a function of distance from a reference pixel at 330, 1384 and 2368~MHz. Although the angular distributions of intensity at 1.4 and 2.4~GHz do not constrain the diffusion of CRs into the dense cores of Sgr~B2, \autoref{fig:lbul} shows that the 330~MHz intensity distribution does constrain the diffusive suppression factor to be $\chi<0.1$ for a $B_\perp=0.8$~mG.  This reinforces the limit found in Figure 10 of \cite{Protheroe2008} where, on the basis of the integrated flux density, it was found that the total flux density from this region implies that for Sgr~B2, $\chi<0.02$. \autoref{eq:diff} and \autoref{fig:lbul} shows that one expects a limb-brightening effect which is determined by the diffusion suppression coefficient (i.e., the lower $\chi$ is, the thinner the brightened limb should be). Thus, since our results show that the coefficient is quite low, we can constrain the width of limb-brightening to be on the order of an arcminute, much broader than the angular resolution of our observations.  

This result shows that the diffusion of CRs is about a few percent of that found in the Galactic disk: a value consistent with that found for the SNR W28, which has been shown to be interacting with nearby molecular clouds \citep{Fujita2009}. 

\subsection{Polarization limits}
We used the noise statistics from the Stokes~V images to place upper limits on the polarized emission from Sgr~B2 of 3.5 and 3~m\jybeam at 1384 and 2368~MHz respectively.  In contrast to the upper limits on the polarized intensity above, the $5\sigma$ RMS error estimate in the total intensity images presented are 50~m\jybeam and 25~m\jybeam at 1384 and 2368~MHz -- a 14 and 8-fold improvement respectively.  

\section{Summary and conclusions}\label{sec:conclusions}
With an RMS sensitivity of 5 m\jybeam and a resolution of $27''\times8''$, we have presented sensitive, high resolution images of Sgr~B region at 2368~GHz.  We have produced 1384~MHz images with the ATCA which match previously published images in sensitivity and resolution at an RMS sensitivity of 10 m\jybeam and a resolution of $47''\times14''$. In summary, our main findings are:

\begin{enumerate}
\item Our observations show that the Sgr~B cloud is dominated by thermal emission, both optically thin and thick.  The 2368~MHz observations reinforce previous findings that the radio continuum emission from Sgr~B2 is dominated by small, dense H{\sc ii} regions, such as Sgr~B2(M) and (N) whose spectral index between 1384 and 2368~MHz are between optically thick and thin thermal emission.  Sgr~B1, on the other hand consists of more diffuse H{\sc ii} regions which have flatter spectral indices between 1384 and 2368~MHz.

\item We found one source of non-thermal emission in the Sgr~B region, Sgr~B2(SC), which has previously been noted as non-thermal.  We have shown that, compared to the 1384 and 2368~MHz observations, the flux density from this source at 330~MHz is significantly absorbed. Using estimates for the molecular density and magnetic field, we find that synchrotron emission from secondary electrons cannot explain the bulk of the non-thermal emission from this object. We show that this is because of the high gas density, which implies that the synchrotron emission from secondary electrons is a function of the {\it volume}, and not proportional to the CR flux and the mass.

\item Using the RMS noise statistics of the Stokes~V emission in Sgr~B2, we have placed limits on the flux density of polarized emission from Sgr~B2 at a level of 3.5 and 3~m\jybeam at 1384 and 2368~MHz respectively. We have used the angular distribution of the total intensity to place limits on the diffusion of the parent CR hadrons into the dense cores of the Sgr~B2 cloud, resulting in an upper limit to the diffusion coefficient of few percent of that found in the Galactic disk.  In light of recent work, this is further evidence that multi-GeV CRs are probably excluded from the dense cores of molecular clouds.

\end{enumerate}
We conclude, then, that with the exception of Sgr~B2(SC) there is (i) no evidence for any non-thermal emission from Sgr~B2 and, therefore, (ii) no evidence of synchrotron emission from secondary electrons in Sgr B2.  The next generation of radio telescopes, such as the EVLA, LWA, ASKAP and SKA, working at lower frequencies may be able to place tighter limits on (or indeed observe) synchrotron emission from secondary electrons due to the increased bandwidth and sensitivity.

\acknowledgments The authors would like to thank the anonymous referee for their comments which greatly improved the manuscript. DIJ would like to thank Jasmina Lazend\'{i}c for help during the preparation of the manuscript. The Australia Telescope Compact Array is part of the Australia Telescope which is funded by the Commonwealth of Australia for operation as a National Facility managed by CSIRO.  This work has made use of NASA's Astrophysics Data System (ADS) abstract service.  This research has made use of the SIMBAD database, operated at CDS, Strasbourg, France.  This research was supported under the Australian Research CouncilÕs Discovery Project funding scheme (project number DP0559991). While this research was conducted Professor R. D. Ekers was the recipient of an Australian Research Council Federation Fellowship (project number FF0345330), J. Ott was the recipient of a Jansky Fellow of the National Radio Astronomy Observatory, and R.M. Crocker was the recipient of the inaugural J.L. William Fellowship.

\end{document}